\documentclass[aps, english, pra, twocolumn, superscriptaddress, nofootinbib, longbibliography]{revtex4-2}
\usepackage{amsmath}
\usepackage[english]{babel}
\usepackage[T1]{fontenc}
\usepackage[utf8]{inputenc}
\usepackage{tikz}
\usetikzlibrary{quantikz2}
\usepackage{pgfplots}
\usepackage{xcolor}
\usepackage{algorithm} 
\usepackage{algpseudocode}
\usepackage[inline]{enumitem}
\usepackage{hyperref}
\usepackage[caption=false]{subfig}
\usepackage{mathtools}
\usepackage{booktabs,tabularx,multirow}
\newcolumntype{Y}{>{\raggedright\arraybackslash}X} 

\definecolor{josblue}{HTML}{4C32FF}
\definecolor{jospurple}{HTML}{B332FF}
\definecolor{jossky}{HTML}{60C0BE}
\DeclarePairedDelimiter{\norm}{\lVert}{\rVert}

\begin{document}
\title{QKD as a Quantum Machine Learning task}
\date{\today}

\author{T. Decker}
\author{M. Gallezot}
\author{S. F. Kerstan}
\author{A. Paesano}

\affiliation{JoS QUANTUM GmbH, c/o Tech Quartier, Platz der Einheit 2, \protect 60327 Frankfurt am Main, Germany \\ \{\emph{thomas.decker, marcelin.gallezot, sven.kerstan, alessio.paesano}\}@jos-quantum.de}

\author{A. Ginter}
\author{W. Wormsbecher}
\affiliation{Bundesdruckerei GmbH, Kommandantenstrasse 18, 10969 Berlin, Germany\\ \{\emph{anke.ginter, wadim.wormsbecher}\}@bdr.de}

\begin{abstract}
We propose considering Quantum Key Distribution (QKD) protocols as a use case for Quantum Machine Learning (QML) algorithms. We define and investigate the QML task of optimizing eavesdropping attacks on the quantum circuit implementation of the BB84 protocol. QKD protocols are well understood and solid security proofs exist enabling an easy evaluation of the QML model performance. The power of easy-to-implement QML techniques is shown by finding the explicit circuit for optimal individual attacks in a noise-free setting. For the noisy setting we find, to the best of our knowledge, a new cloning algorithm, which can outperform known cloning methods.  
Finally, we present a QML construction of a collective attack by using classical information from QKD post-processing within the QML algorithm.

\end{abstract}


\maketitle

\section{Introduction}
Quantum Machine Learning (QML) \cite{qml1, qml2} has attracted significant interest in recent years with the promise of improving our capabilities of learning from data using the computing potential of quantum systems. Because of the state of quantum hardware, it is currently not possible to run QML algorithms on noise-free large-scale systems. For this reason, quantum algorithms are mostly benchmarked using numerical simulations. This limits QML use cases to relatively small-scale models on reduced data sets and makes the comparison to classical models a challenging task. For a comprehensive analysis of the current challenges of benchmarking QML algorithms, we refer readers to \cite{qmlReview}.

In the light of this situation, we see Quantum Key Distribution (QKD) as an attractive field of research for QML: QKD protocols operate on small-scale quantum systems, and those systems are already sufficient for interesting QML applications, as we will show in this work.

The literature on quantum key distribution (QKD) has grown from a few pioneering papers in the 1980s, such as \cite{BB84}, which established the idea of QKD with the introduction of the BB84 protocol, to an impressive number of new protocols, for example the prominent E91 \cite{E91} and six-state \cite{sixstate} 
protocols, security proofs for many protocols, e.g. \cite{Shor_2000} for BB84, and later also security proofs that deal with finite-size keys \cite{renner2008security}.
An important part of QKD is the classical post-processing, which consists of error reconciliation \cite{Cascade, LDPC} and privacy amplification \cite{PrivacyAmplification}.
We recommend the review \cite{bigReview} for readers seeking a detailed and accessible overview over the QKD literature.

In this paper, we show that using a sub-discipline of QML called Quantum Circuit Learning (QCL) \cite{QCL}, it is possible to find optimal attacks on the BB84 protocol. QCL is an hybrid quantum-classical algorithm where the weights of a parametrized quantum circuit are classically optimized to perform a given task by minimizing a chosen loss function (sometimes also known as as target objective or learning objective). QKD protocols and attacks on them can be expressed through parametrized quantum circuits. With a proper choice of loss function, QCL can then be used to optimize such an attack.

We show that with a relatively shallow QCL ansatz and appropriate loss functions, we find the optimal individual attack on the BB84 protocol with a noise-free quantum channel, which is known to be the Phase Covariant Cloning Machine (PCCM)~\cite{PCCM}. That this is possible was noticed earlier in \cite{VQC}. We also show that in the presence of noise, a QCL attack can outperform the PCCM. This demonstrates that it is not the optimal individual attack in general. In the final example we apply QCL to construct a, within the expected accuracy, optimal\footnote{Strictly speaking, we optimized only the measurement, not the interaction with Alice and Bob's quantum channel. The measurement we find with QCL is optimal.} collective attack on BB84 when parity information about pairs of transmitted qubits is exchanged between Alice and Bob for post-processing the key. 

We stress that our examples do not break any existing BB84 security proofs. From a QKD point of view, we provide a convenient construction of explicit representations of attacks on the protocol\footnote{Note that we do not consider attacks which exploit imperfect physical implementations (i.e. side channels) of the protocol, but attacks on the (perfectly implemented) protocol itself.}. Some explicit constructions exist in the literature, such as in \cite{PCCM} and \cite{molotkov2007explicit}, but in the light of existing security proofs, the actual attacks appear to be of limited interest. However, from a QML point of view, our examples demonstrate the surprising power of relatively simple QCL constructions and thus a guideline for circuit design and evaluation.

\section{Brief review of the BB84 protocol} \label{BB84_section}
The BB84 protocol was originally suggested in 1984 by Bennett and Brassard. It allows two parties, Alice and Bob, to generate and distribute a classical binary key over a quantum channel. Additionally, they require a verified and authenticated classical channel. Even in the presence of an eavesdropper Eve, the security of the transmission can be guaranteed in a well-defined range of circumstances based on the laws of physics (e.g. no-cloning theorem) and appropriate post processing. The $n$-th bit of the secret key is established with the following steps:

\begin{enumerate}
    \item \textbf{State preparation}: Alice randomly chooses a bit value $x\in\{0,1\}$ followed by another random bit $b_A\in\{0,1\}$ that determines one of two orthogonal bases -- the $Z$-basis ($b_A = 0$) or $X$-basis ($b_A = 1$) -- used to encode the bit $x$ as a quantum state. For $x=0$, the prepared qubit is $\ket{0}$ for $b_A=0$ and $\ket{+}$ for $b_A=1$. For $x=1$, the prepared qubit is $\ket{1}$ for $b_A=0$ and $\ket{-}$ for $b_A=1$.
    \item \textbf{Transmission}: The quantum state is sent through the quantum channel from Alice to Bob.
    \item \textbf{Measuring}: Bob randomly selects 
    $b_B\in\{0,1\}$ that determines one of the two orthogonal bases $Z$ ($b_B=0$) or $X$ ($b_B=1$) to measure the received quantum state and store the binary result.
    \item \textbf{Sifting}: Alice and Bob reveal their basis choices over the classical channel. If the basis coincides, i.e. $b_A = b_B$, the measured bit is used as the $n$-th bit of their raw key. Otherwise, they discard the bit and start over with the process for the $n$-th bit. 
\end{enumerate}
Note that the sifting step can be performed after the transmission of each individual bit or just once for an entire block of bits. 

With an ideal quantum channel, Alice and Bob would always end up with identical raw keys. However, in a realistic scenario, errors occur in the quantum channel, either due to environmental noise or the actions of an eavesdropper, causing some bits in Bob's key to differ from Alice's.
Therefore, after the transmission of the quantum states, Alice and Bob have to apply additional post-processing steps to obtain identical and secure keys. First, Alice and Bob will communicate classical information over the verified channel in order to detect and correct any errors between their raw keys obtaining an error-free reconciled key. This can be achieved with a variety of error reconciliation protocols, such as, for example Cascade \cite{Cascade} or protocols based on Low-Density Parity-Check (LDPC) \cite{LDPC} codes. This will necessarily reveal information about the raw key and we show in section~\ref{QCLPCCMCOLLECTIVE} how it can be used by an eavesdropper. A necessary condition for a secure key is that an attacker cannot obtain the reconciled key in full.  Privacy amplification aims at reducing the information obtained by an eavesdropper by distilling the reconciled key. However, privacy amplification will not play a role in our consideration.

\section{Quantum Circuit Learning and Attacks on the BB84 protocol} \label{QCL}
In our examples, we deal with two types of attacks: \textit{individual attacks}, in which Eve uses the same quantum circuit to interact with the quantum channel for every qubit that is transmitted. Further, for each such qubit transmission, the measurement she applies to her quantum system is independent of all other qubit transmissions. In \textit{collective attacks}, Eve still applies the same quantum circuit for each qubit, but the quantum system she measures in the end may now depend on all the qubits that were transmitted. Collective attacks are generally more powerful than individual attacks, as our example in section \ref{QCLPCCMCOLLECTIVE} illustrates.

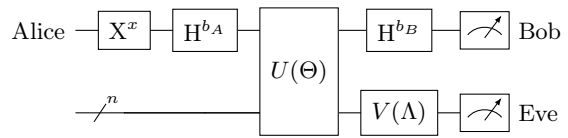
\begin{figure}[!ht]
\begin{center}
\begin{quantikz}[thin lines, classical gap=0.09cm, column sep = 0.3cm]
\lstick{Alice}& \gate{\text{X}^{x}} & \gate{\text{H}^{b_A}} & \gate[2]{U(\Theta)} & \gate{\text{H}^{b_B}} & \meter{} \rstick{Bob}\\
\lstick{}& \qwbundle{n} & \qw & \qw & \gate{V(\Lambda)}& \meter{} \rstick{Eve}
\end{quantikz}
\end{center}
\caption{Individual attack on BB84 - $x\in\{0, 1\}$ is the bit sent by Alice in the $Z$-basis ($b_A = 0$) or the $X$-basis ($b_A = 1$). Similarly, Bob can measure in the $Z$-basis ($b_B = 0$) or the $X$-basis ($b_B = 1$).}
\label{fig:circuit}
\end{figure}

We model an individual attack on the BB84 protocol with a quantum circuit as shown in figure~\ref{fig:circuit}. The communication channel is represented by a single qubit with the two parties Alice and Bob at each end. Eve has access to an ancillary quantum register and can interact with the QKD line (we assume that this process is error-free) during the transmission using the unitary operation $U(\Theta)$. With access to quantum memory, Eve may also delay another unitary operation $V(\Lambda)$ and measurement until further classical information is broadcasted by Alice and Bob. Eve can then perform operations and measurements on the ancilla system that make use of this information. For example, in an individual attack such as the ones described in section~\ref{QCLPCCM} and~\ref{QCLPCCMNOISE}, Eve uses two sets of weights for $V(\Lambda)$ depending on the basis that was used by Alice and Bob. We extend this approach to collective attacks in section~\ref{QCLPCCMCOLLECTIVE}.

A full analysis includes eight different quantum circuits corresponding to all possible choices in state preparation, i.e. ${\ket{0}, \ket{1}, \ket{+}, \ket{-}}$, and Bob's choice of measurement basis. For each configuration we compute the density matrix of the system. Bob's fidelity is defined as the probability to measure Alice's bit choice. This equals the quantum fidelity 
    \begin{equation}
        \label{eq:quantum_fid}
        F(\rho_A, \rho_B) = \left(\operatorname{Tr} \sqrt{\sqrt{\rho_A} \rho_B \sqrt{\rho_A}}\right)^2
    \end{equation}
    where $\rho_A$ is Alice's density matrix and $\rho_B$ represents Bob's density matrix after Eve's attack and before his measurement basis choice.

Since only cases with matching bases are sifted, we restrict our analysis to cases where Alice and Bob have chosen the same basis. As the four different configurations ${\ket{0}, \ket{1}, \ket{+}, \ket{-}}$ are equally likely to happen, we define $F_{AB}(\Theta, \Lambda)$ as the average fidelity over the four possible cases. The average error rate observed by Alice and Bob can then be written $\varepsilon = 1 - F_{AB}$. Similarly, we define Eve's fidelity as $F_{AE}(\Theta, \Lambda)$. 

\begin{figure}
\centering
\subfloat[][]{
	\begin{tikzpicture}
            \node[scale=0.73]{
            \begin{quantikz}[thin lines, classical gap=0.09cm, column sep = 0.3cm]
    		& \gate{R_X} & \gate{R_Y} & \gate{R_Z} & \ctrl{1} & \\
    		& \gate{R_X} & \gate{R_Y} & \gate{R_Z} & \targ{} &
    	\end{quantikz}
        };
        \end{tikzpicture}
        \label{fig:ansatz2qubits}
}
\qquad
\subfloat[][]{
        \begin{tikzpicture}
            \node[scale=0.73]{
            \begin{quantikz}[thin lines, classical gap=0.09cm, column sep = 0.3cm]
    		& \gate{R_X} & \gate{R_Y} & \gate{R_Z} & \ctrl{1} & \qw & \targ{} & \\
    		& \gate{R_X} & \gate{R_Y} & \gate{R_Z} & \targ{} & \ctrl{1} & \qw & \\
    		& \gate{R_X} & \gate{R_Y} & \gate{R_Z} & \qw & \targ{} & \ctrl{-2} &
    	\end{quantikz}
        };
        \end{tikzpicture}
}
    \caption{Example of one layer of QCL ansatz for two and three qubits. Each single-qubit rotation is associated with a trainable parameter.}
    \label{fig:ansatz}
\end{figure}
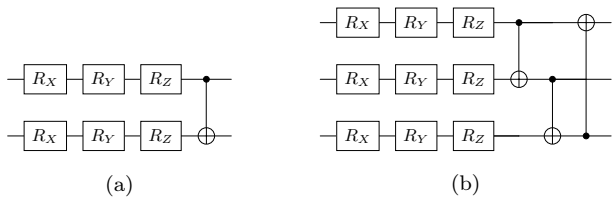

These values define the properties of the attack and can be optimized with QCL using the unitaries $U(\Theta)$ and $V(\Lambda)$ as parametrized quantum circuits. To do so, we define an appropriate loss ${\cal L}(\Theta, \Lambda)$ as a function of Bob and Eve’s fidelities. We minimize it with gradient-based methods over a fixed number of steps by optimizing the weights of the parametrized unitaries. We choose a circuit ansatz such as in figure~\ref{fig:ansatz} for $U(\Theta)$ and $V(\Lambda)$ consisting of alternating layers of single-qubit rotations and entangling CNOT gates arranged in a ring topology. Each single qubit gate carries a trainable parameter. This architecture is used in many QML applications \cite{VQA} and referred to as Hardware-Efficient Ansatz (HEA) \cite{HEA, HEA_schuld}.

\section{Optimal individual attacks on BB84 from QCL}\label{QCLPCCM}
\begin{figure}[!ht]
    \begin{center}
        \begin{tikzpicture}
            \node[scale=0.85]{
            \begin{quantikz}[thin lines, column sep = 0.3cm]
            \lstick{Alice}&\gate{R_X(\frac{\pi}{2})}&\ctrl{1} &\gate{R_Y(-\pi)}&\gate{R_X(-\frac{\pi}{2})} \slice[style=black]{} & & \rstick{Bob}  \\
            &\qw &\gate{R_Y(\theta)}&\ctrl{-1} &\gate{R_X(-\frac{\pi}{2})} & \gate{H^{b_E}} & \rstick{Eve}
            \end{quantikz}
            };
        \end{tikzpicture}
    \end{center}
    \caption{Quantum circuit for the PCCM with angle $\theta$. The additional H gate is delayed by Eve until the basis reveal happens and only applied if Alice and Bob have used the $X$-basis.}
    \label{fig:pccm}
\end{figure}
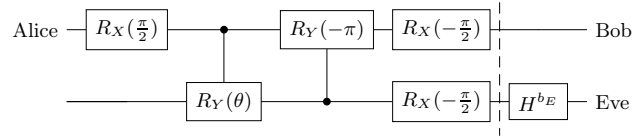

The optimal individual attack on BB84 is given by the Phase Covariant Cloning Machine (PCCM) in the case of an error-free transmission channel \cite{PCCM}. A quantum circuit for implementing the PCCM is shown in figure~\ref{fig:pccm}. The fidelities for Bob and Eve are functions of a single rotation angle $\theta$:
\begin{subequations}
    \begin{align}
&F_{AB} = \frac{1 + \cos\theta/2}{2} \label{eq:pccm_fidelity_alice_bob}\\
&F_{AE} = \frac{1 + \sin\theta/2}{2} \label{eq:pccm_fidelity_alice_eve}
    \end{align}    
\end{subequations}
The symmetric case $F_{AB} = F_{AE} \approx 0.853$ is given by $\theta = \pi/2$.

This optimal individual attack is easily reproduced using QCL.
With a single ancillary qubit for Eve, we implement $U(\Theta)$ with two layers of the circuit shown in figure ~\ref{fig:ansatz2qubits} resulting in a total of 12 parametrized rotations and 2 entangling CNOT gates. For $V(\Lambda)$ we use only 3 parametrized rotations and two sets of weights, one for when Alice and Bob used the $Z$-basis, the other for the $X$-basis. This brings the total of trainable parameters in the circuit to 18. For the optimization of the attack we use the following loss function:
\begin{equation}
{\cal L}(\Theta, \Lambda)=\alpha \left( F_{AB}(\Theta, \Lambda) - f \right)^2 - F_{AE}(\Theta, \Lambda)
\end{equation}
where $\alpha$ is a weighting parameter and $f$ is the target value for $F_{AB}(\Theta, \Lambda)$. With the first term we can force the optimization to target a specific fidelity for Bob while the second term is used to maximize Eve's fidelity. 

\algrenewcommand\algorithmicrequire{\textbf{Parameters:}}
\begin{algorithm}[H]   
    \caption{QKD attack optimization with QCL} 
    \begin{algorithmic}
        \Require $f$ target fidelity for Bob, $\alpha$ weighting parameter
        
        \State \textbf{Initialize} weights $\Theta$, $\Lambda$ from a normal distribution
        \For {$step=1,\ldots,n_{steps}$}
            \State \textbf{Simulate} density matrices $\{\rho(x, b_A=b_B, \Theta, \Lambda)\}$ \Comment{One density matrix per combination of transmitted state and chosen basis (4 in total)}
            \State \textbf{Compute} average fidelities $F_{AB}$ and $F_{AE}$
            \State \textbf{Compute} loss function $\mathcal{L}(\Theta, \Lambda)=\alpha \left(F_{AB} - f\right)^2 - F_{AE}$
            \State \textbf{Compute} gradients $\nabla_\Theta\mathcal{L}(\Theta, \Lambda)$ and $\nabla_\Theta\mathcal{L}(\Theta, \Lambda)$ using the parameter-shift rule
            \State \textbf{Update} $\Theta$ and $\Lambda$ with the chosen optimizer
        \EndFor
    \end{algorithmic} 
    \label{algorithm:QCL_training}
\end{algorithm}

For each choice of hyperparameters $\alpha$ and $f$, a gradient descent optimization is conducted on the parameters of the circuit ansatz according to algorithm \ref{algorithm:QCL_training}. We use the Adam optimizer \cite{AdamOptimizer} to minimize the loss function $\cal L$ over 100 iterations with a learning rate of $0.1$. For the target fidelity we used random values in $[0.5, 1]$ and the weight $\alpha$ was heuristically set to 10. For the simulation of the circuit and for the optimization we use a simulator for mixed-states from the Pennylane framework~\cite{Pennylane}.

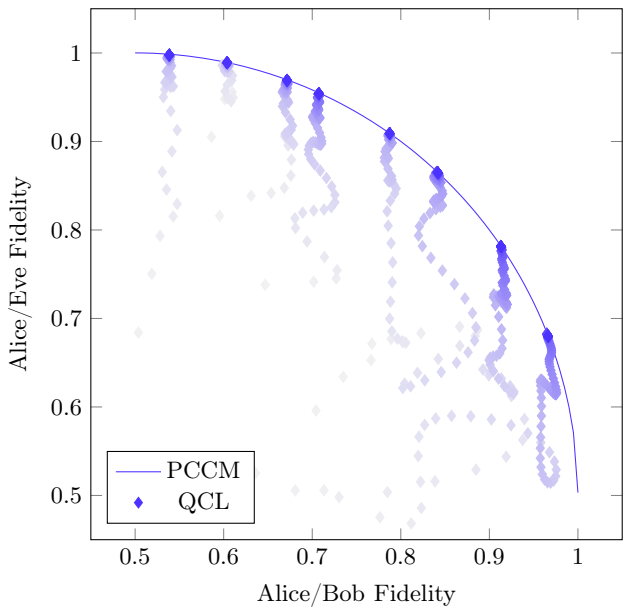
\begin{figure}[!ht]
\begin{tikzpicture}
\begin{axis}[
	width=\columnwidth,
	height=\columnwidth,
    xlabel={Alice/Bob Fidelity},
    ylabel={Alice/Eve Fidelity},
    xlabel style={at={(axis description cs:0.5,+0.01)},anchor=north},
    ylabel style={at={(axis description cs:+0.08,0.5)},anchor=south},
    legend pos=south west,
	point meta min=0,
    point meta max=99,
    colormap name=custom,
    xmin = 0.5,
    xmax = 1,
    ymin = 0.5,
    ymax = 1,
    enlarge x limits={0.1}, 
    enlarge y limits={0.1}, 
]
\pgfplotsset{
    colormap={custom}{
        rgb255(0pt)=(241,241,241);
        rgb255(100pt)=(76,50,255)
    },
}
\addlegendimage{line legend, color=josblue, mark=none};
\addlegendentry{PCCM};
\addlegendimage{only marks, color=josblue, mark=diamond*};
\addlegendentry{QCL};
\addplot [
    domain=0.5:1, 
    samples=100, 
    color=josblue,
    ]
    {0.5*(2*(x*(1-x))^0.5+1)};
\addplot+[scatter, 
    only marks, 
    mark=diamond*, 
    scatter src=explicit,
    scatter/use mapped color={draw=mapped color, fill=mapped color},
    visualization depends on={\thisrow{step} \as \stepValue},
    line width=0.5pt,
    mark options={scale=1.0}] table[x=f_ab, 
    y=f_ae, 
    meta=step, 
    col sep=comma] {test_gradient_plot.csv};
\end{axis}
\end{tikzpicture}
\caption{Comparison of the PCCM with the results of QCL for an attack on an individual qubit. The blue line represents the fidelities of a PCCM with different rotation angles $\theta$. We conducted 8 QCL rounds with varying target fidelities for Bob. Optimization results are marked by diamonds, colored according to the training step.}
\label{fig: pccm vs qcl}
\end{figure}

The results are shown in figure \ref{fig: pccm vs qcl}. We performed 8 training rounds, each one for a different target fidelity $f$ for Bob. For each round, we display the intermediate results during the training phase in light colors, while the final result of the training is dark. We conclude that QCL converges to the same fidelities as the PCCM thus resulting in an optimal individual attack on BB84 with little effort.

\section{QCL attacks on BB84 with noisy quantum channels} \label{QCLPCCMNOISE}
We now assume the presence of noise in the quantum channel which is not caused by Eve. The noise can occur between Alice and Eve or Eve and Bob. It can be caused by environmental factors or have directly been introduced by Alice and Bob to increase the robustness of BB84 against attacks \cite{infoSecurityProof}. We assume that Eve was not able to eliminate the noise. For this analysis, we choose different error models that are readily available in the Pennylane framework like bit- or phase-flip errors, amplitude- or phase-damping errors. We consider different error strengths.

\begin{figure}[!ht]
\begin{tikzpicture}
\begin{axis}[
	width=\columnwidth,
	height=\columnwidth,
    xlabel={Alice/Bob Fidelity},
    ylabel={Alice/Eve Fidelity},
    xlabel style={at={(axis description cs:0.5,+0.01)},anchor=north},
    ylabel style={at={(axis description cs:+0.08,0.5)},anchor=south},
    legend style={
            at={(0.5,-0.15)},
            anchor=north,
            legend columns=3,
            /tikz/every even column/.append style={column sep=0.1cm}
        },
    legend cell align={left},
    enlarge x limits={0.05}, 
    enlarge y limits={0.05}, 
]

    \addlegendimage{color=jossky, mark=none};
    \addlegendentry{$Z$-basis};
    \addlegendimage{color=jospurple, mark=none};
    \addlegendentry{$X$-basis};
    \addlegendimage{color=josblue, mark=none};
    \addlegendentry{Average};
    \addlegendimage{only marks, color=black, mark=diamond*};
    \addlegendentry{QCL};
    \addlegendimage{line legend, color=black, mark=none};
    \addlegendentry{PCCM};
    
    \addplot+[only marks, mark=diamond*, color=jossky, mark options={fill=jossky}] table[x=f_ab_Z,y=f_ae_Z, col sep=comma] {cloner_noise.csv};
    \addplot+[only marks, mark=diamond*, color=jospurple, mark options={fill=jospurple}] table[x=f_ab_X,y=f_ae_X, col sep=comma] {cloner_noise.csv};
    \addplot+[only marks, mark=diamond*, color=josblue, mark options={fill=josblue}] table[x=f_ab,y=f_ae, col sep=comma] {cloner_noise.csv};
    \addplot [
        domain=0.5:1, 
        samples=100, 
        color=jospurple,
        ]
        {(0.5^2 - (x-0.5)^2)^0.5+0.5)};
    \addplot [
        domain=0.5:0.75, 
        samples=100, 
        color=jossky,
        ]
        {(0.25^2 - (x-0.5)^2)^0.5+0.5)};
    \addplot [
        domain=0.5:0.875, 
        samples=100, 
        color=josblue,
        ]
        {(0.375^2 - (x-0.5)^2)^0.5+0.5)};
    \addplot+[mark=None, color=jossky, style=dashed] table[x=f_ab_Z,y=f_ae_Z, col sep=comma] {cloner_theory.csv};
    \addplot+[mark=None, color=jospurple, style=dashed] table[x=f_ab_X,y=f_ae_X, col sep=comma] {cloner_theory.csv};
    \addplot+[mark=None, color=josblue, style=dashed] table[x=f_ab,y=f_ae, col sep=comma] {cloner_theory.csv};
\end{axis}
\end{tikzpicture}
\caption{PCCM and QCL results for a communication channel with 25\% chance of a bit-flip error before the attack. The cloning performance is different for the two bases of the BB84 protocol.}
\label{fig:noise}
\end{figure}
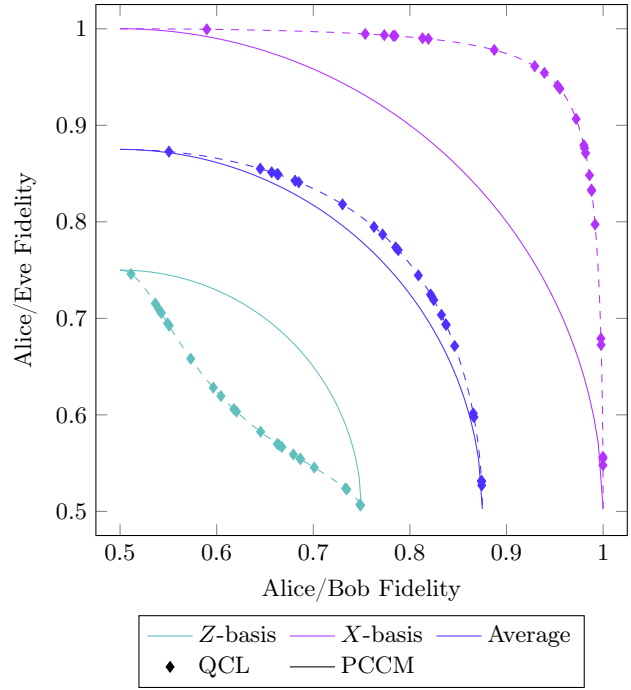

For each scenario, we conduct 25 different QCL training rounds utilizing identical ansätze, loss functions and parameters as detailed in section \ref{QCLPCCM}. As an example, we present details of a scenario where a bit-flip error has a 25\% chance of occurring prior to Eve's attack. The results of the QCL optimization are compared to the fidelities obtained with the PCCM in that scenario in figure~\ref{fig:noise}. The final step of the optimization is displayed and marked by a diamond symbol.

Considering the average fidelity over both possible bases of the BB84 protocol, we observe that QCL generally slightly outperforms the PCCM. Note that bit-flip errors affect the two bases of the BB84 protocol differently. While inducing flips from \ket{0} to \ket{1} and from \ket{1} to \ket{0}, it does not alter the states of the $X$-basis up to a phase factor that does not affect Bob's measurement. QCL is able to beat the PCCM by sacrificing fidelity in the $Z$-basis for an increased fidelity in the unaffected $X$-basis.

The found solution is an example of what we refer to as \textit{imbalanced cloner}, because the cloning fidelity depends on the basis. Note that the imbalanced cloner will perform worse than a PCCM in the absence of noise. 

For more general error models, we found that when the errors affect the two bases of the BB84 protocol differently, a QCL ansatz beats the PCCM which therefore is not the optimal individual attack anymore.

\begin{figure}[!ht]
\begin{center}
\begin{tikzpicture}
\node[scale=0.7]{
\begin{quantikz}[thin lines, column sep = 0.3cm]
& \gate{R_X(-\frac{\pi}{2})} & & & \ctrl{1} & \gate{R_X(\frac{\pi}{2})} & \ctrl{1} & \slice[style=black]{} &&\\

& \gate{R_X(-\frac{\pi}{2})} & \gate{R_Z(\pi)} & \gate{R_Z(\psi)}& \targ{} & \gate{R_Z(\phi)} & \targ{} & \gate{R_X(-\frac{\pi}{2})} & \gate{R_X(\frac{3\pi}{2})^{b_E}}&
\end{quantikz}
};
\end{tikzpicture}
\end{center}
\caption{Quantum circuit for the imbalanced cloner with two parameters $\psi$ and $\phi$.}
\label{fig:imbcloner}
\end{figure}
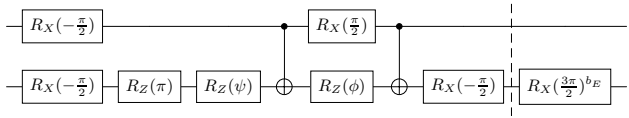

We will now further investigate how the QCL solution is able to beat the PCCM in such scenarios.
For notational convenience, we define the centered fidelity $C_{AB, Z} \coloneq 2 F_{AB, Z} - 1$, where $F_{AB, Z}$ denotes the average fidelity of Bob when Alice uses the $Z$-basis (eq. \ref{eq:quantum_fid}). We characterize the effects of noise using parameters $\alpha$, $\beta$, $\gamma$, $\delta \in [0, 1]$ such that the centered fidelity between Alice and Bob in the $Z$-basis $C_{AB, Z}$ becomes $\Tilde{C}_{AB,Z}
 = \alpha C_{AB, Z}$, while simultaneously:
\begin{align}
C_{AE, Z} &\rightarrow \Tilde{C}_{AE, Z} = \gamma \, C_{AE, Z} \\
C_{AB, X} &\rightarrow \Tilde{C}_{AB, X} = \beta \, C_{AB, X} \\ 
C_{AE, X} &\rightarrow \Tilde{C}_{AE, X} =  \delta \, C_{AE, X}
\end{align}
After verifying that increasing the size of the HEA described in sections \ref{QCL} and \ref{QCLPCCMNOISE} to more qubits and more layers does not further improve the performance of the QCL solution, we tried to simplify the quantum circuit as much as possible. We removed quantum gates that did not improve the performance, and we reduced the number of qubits. The result of those efforts is the much simpler circuit for the imbalanced cloner in figure \ref{fig:imbcloner}. This cloner has two parameters $\psi$ and $\phi$ and using equation \ref{eq:quantum_fid}, we can calculate the centered fidelities $C_{AB, Z} = \sin{\psi}$, $C_{AB, X} = \cos{\phi}$, $C_{AE, Z} = -\sin{\phi}$   and   $C_{AE, X} = \cos{\psi}$. This circuit outperforms the PCCM when $\alpha\gamma \neq \beta\delta$. The $R_X(3\pi/2)$ gate is only applied if Alice and Bob have used the $X$-basis.

As opposed to the PPCM, which has only one free parameter, the imbalanced cloner has two free parameters, $\psi$ and $\phi$. This enables it to adapt to two degrees of freedom: the fidelity between Alice and Bob and the particular error model at at hand. Given a fixed value of $\psi$, we determine the optimal choice for $\phi$ as a function of the noise characteristics $\alpha, \beta, \gamma, \delta$. To maximize $\Tilde{C}_{AB} = \frac{1}{2}(\Tilde{C}_{AB, Z} + \Tilde{C}_{AB, X})$ and $\Tilde{C}_{AE} = \frac{1}{2}(\Tilde{C}_{AE, Z} + \Tilde{C}_{AE, X})$, we solve for a vanishing determinant of the Jacobian:

\begin{equation}
\det\begin{vmatrix}
\frac{\partial \Tilde{C}_{AB}}{\partial \psi} & \frac{\partial \Tilde{C}_{AB}}{\partial \phi} \\
\frac{\partial \Tilde{C}_{AE}}{\partial \psi} & \frac{\partial \Tilde{C}_{AE}}{\partial \phi}
\end{vmatrix}
= \frac{1}{4}\det\begin{vmatrix}
\alpha  \cos\psi & -\beta  \sin\phi \\
-\delta  \sin\psi & -\gamma \cos\phi
\end{vmatrix}
\end{equation}

This yields the relationship:
\begin{equation}
\phi = - \arctan{\left(\frac{\alpha\gamma}{\beta\delta} \cdot \frac{\cos\psi}{\sin\psi}\right)}
\end{equation}

Using this result, one can find out the maximal value of $\Tilde{C}_{AE, Z}$ (resp. $\Tilde{C}_{AE, X}$) as a function of $\Tilde{C}_{AB, Z}$ (resp. $\Tilde{C}_{AB, X}$). 

\begin{equation}
\Tilde{C}_{AE, Z} = \frac{\gamma}{\sqrt{1 + \left(\frac{\beta\delta}{\alpha\gamma}\right)^2\dfrac{\Tilde{C}_{AB, Z}^2}{\alpha^2 - \Tilde{C}_{AB, Z}^2}}}
\end{equation}

\begin{equation}
\Tilde{C}_{AE, X} = \frac{\delta}{\sqrt{1 + \left(\frac{\alpha\gamma}{\beta\delta}\right)^2\dfrac{\Tilde{C}_{AB, X}^2}{\beta^2 - \Tilde{C}_{AB, X}^2}}}
\end{equation}

Note that when $\alpha = \beta$ and $\gamma = \delta$, one obtains the equations of the PCCM and a quantum circuit equivalent to that of the PCCM in figure \ref{fig:pccm}. The imbalanced cloner circuit in figure \ref{fig:imbcloner} can thus be understood as a generalization of the PCCM with a new degree of freedom that allows the cloner to adapt to the imbalance between the two BB84 bases.

\section{Collective attacks on BB84} \label{QCLPCCMCOLLECTIVE}
For an eavesdropper, individual attacks on BB84 are not optimal \cite{infoSecurityProof, molotkov2007explicit} and so we make a simple ansatz for a stronger attack as follows: Eve uses a PCCM on each qubit that Alice sends to Bob and stores her approximate copies in her quantum memory. Following the transmission, Eve performs a measurement on the whole register. This type of attack is referred to as collective. The classical information exchanged  by Alice and Bob during error reconciliation can be used by Eve for optimizing the measurement. Here, we consider an example were the Cascade protocol is used~\cite{Cascade}. The Cascade algorithm works by exchanging parities of blocks of the raw key in order to detect and correct errors in it.

What follows can be viewed as one of two scenarios: In the first scenario, we consider a toy model, in which Alice and Bob use BB84 to generate a key with a length of just 2 bits. In this case, the Cascade algorithm collapses to the transmission of a single bit of parity information about that pair. If the parity does not match between Alice and Bob, Cascade would reveal the full key in this example, and the key therefore is abandoned.  In the second scenario, Alice and Bob generate a much longer key, and Eve uses the parity information generated by Cascade to identify pairs and larger tuples of key bits, for each of which Eve tries to use the parity extracted from Cascade to perform the optimal measurement. Our example here is restricted to the parities of pairs of bits. For larger tuples Eve could either drop the extra information (which is the worst choice, of course), use a pretty good measurement \cite{PGM}, or use QCL to learn the best possible attack, just like in our two qubit example here.\\
The construction of the collective attack (or part of the collective attack, see above) reduces to the following problem: Eve has two PCCM copies of two BB84 states that Alice sent and she knows the parity of those states. What is the optimal measurement for Eve to distinguish either 00 from 11 in the case of even parity or 01 from 10 in the case of odd parity?
As an example, we investigate the case where the average fidelity $F_{AB}$ that Alice and Bob measure for their channel is $0.892$.\footnote{This number is chosen such that the widely known lower bound for the Quantum  Bit Error Rate (QBER) of $11\%$ is realized.
}\\
Given the values $a_0 a_1$ of the two bits Alice sent, the values $b_0 b_1$ that Bob measured and the values $e_0 e_1$ that Eve obtained from her attack, we define the success probability $P_E$ for Eve by $P_E= \frac{P(a_0 a_1 = b_0 b_1 = e_0 e_1)}{P(a_0 a_1 = b_0 b_1)}$. Events where $a_0 a_1 \neq b_0 b_1$ are excluded from the statistics\footnote{We assume that only Bob uses the key for encryption, so Eve will only benefit from cloning the transmission successfully if Bob's bit values also agree with Alice's.}.
We begin by examining individual attacks on the two-qubit setup. Eve can achieve the optimal result by measuring only the first qubit, and deducing the value of the second qubit using the parity. According to equations \ref{eq:pccm_fidelity_alice_bob} and \ref{eq:pccm_fidelity_alice_eve}, Eve measures the qubit with a fidelity of $F_{AE} \approx 0.810$, which then also is the probability for getting both bits correct. 

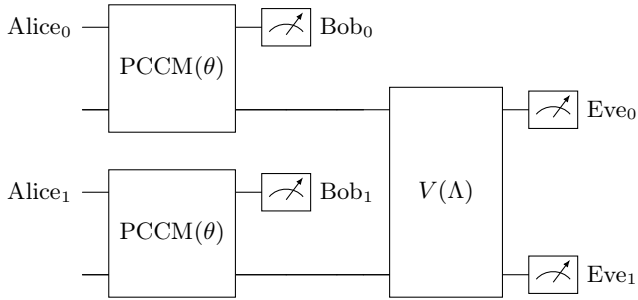
\begin{figure}[!htb]
    \begin{center}
    \begin{quantikz}[thin lines, column sep = 0.35cm, slice style = blue]
    \lstick{$\text{Alice}_0$}& \gate[2][1.5cm]{\text{PCCM}(\theta)} & \meter{} \rstick{$\text{Bob}_0$}\\
    & \qw & \qw & \qw & \qw & \gate[3][1.5cm]{V(\Lambda)} & \meter{} \rstick{$\text{Eve}_0$}\\
    \lstick{$\text{Alice}_1$}& \gate[2]{\text{PCCM}(\theta)} & \meter{} \rstick{$\text{Bob}_1$}\\
    & \qw & \qw & \qw & \qw & \qw & \meter{} \rstick{$\text{Eve}_1$}
    \end{quantikz}
    \end{center}
    \caption{Quantum circuit for the collective attack on two qubits. Eve uses a PCCM attack with the same angle $\theta$ on both qubits. She then delays another unitary $V$ that depends on the bases that were used by Alice and Bob as well as the parity information.}
    \label{fig:collective}
    \end{figure}

To try and beat the result from the individual attack, we again use QCL to optimize Eve's attack.
We take one intermediate step: we simplify the output of Eve's two PCCMs which she has in her quantum memory. Eve knows the basis of the states Alice sent and applies the following operations on each of her approximate copies:
\begin{itemize}
\item $Z$-basis: Eve applies an $S$ gate and then a $Z$ gate
\item $X$-basis: Eve applies an $H$ gate and then an $S$ gate
\end{itemize}
Then, regardless of the basis Alice chose for the transmission, Eve's approximate copies of "0" or "1" sent by Alice are the density matrices $\rho_0$ or $\rho_1$, given by
$$
\rho_0=\left(\begin{array}{cc}0.810&0.392\\0.392&0.190\end{array}\right) \quad {\rm and} 
\quad \rho_1=\left(\begin{array}{cc}0.190&0.392\\0.392&0.810\end{array}\right)
$$

Note that these density matrices are not obtained from the
PCCM system by tracing out Bob. Rather, they describe the system that Eve obtains when Bob has correctly measured the state that Alice prepared.

Now Eve uses an HEA for $V(\Lambda)$ with two qubits (i.e. there are no ancillary qubits) and two layers for a total of 12 parametrized rotations and two entangling CNOT gates. We use two distinct sets of weights depending on the parity transmitted by Alice, leading to a total of 24 trainable parameters. This setup is shown in figure \ref{fig:collective}. Since Eve wants to distinguish two states for a given parity, she only measures one of her two qubits, without loss of generality, the first one. In a case where the parity is 0, a measurement result of 0 indicates 00 as the state Alice sent, while measuring 1 indicates 11, and similarly for the case where the parity is 1. The loss function is then given by the (negative) success probability $-P_E=-\frac{P(a_0 a_1 = b_0 b_1 = e_0 e_1)}{P(a_0 a_1 = b_0 b_1)}$, which we obtain from Eve's density matrix in the simulation. As in the earlier examples, the training is done with 100 iterations of the Adam optimizer and a learning rate of 0.1. We find that the resulting, optimized circuit obtains the correct result for both bits in $89.4\%$ of the cases. As expected, this is significantly better than the $81.0\%$ Eve can obtain from the individual attack.

Finally, we compare this to the optimal solution for distinguishing Eve's two quantum states, which is known to be given by the Helstrom measurement \cite{helstrom1969quantum}. A brief calculation yields 
\begin{equation}
0.5 + 0.5 \, \norm{0.5 \,\rho_0 \otimes \rho_0  - 0.5 \, \rho_1\otimes \rho_1}_1 =89.4\%
\end{equation}
as the optimal success rate for Eve to distinguish the states 00 and 11. The same result holds for odd parity, that is 10 and 01. Note that the norm $\norm{\cdot}_1$ is the trace norm, with $\norm{M}_1 = \operatorname{Tr}\sqrt{M^*M}$.

So again, our approach for constructing Eve's attack with the HEA and QCL gave the optimal result. 

\section{Conclusion and Outlook}

We have shown that QCL methods can be seen as a powerful tool to optimize attacks on the BB84 protocol. Besides the optimization, these methods also provide us with explicit implementations of the attacks as quantum circuits thus introducing a quantum algorithmic treatment of QKD. We have analyzed the results of QCL for three different scenarios: 
\begin{enumerate*}[label=(\arabic*)]
  \item attacks on single qubits,
  \item attacks on single qubits with a noisy transmission line,
  \item collective attacks on two qubits.
\end{enumerate*}

While the result of our first scenario is well known, i.e. a PCCM is the optimal individual attack~\cite{PCCM}, the second scenario leads to a generalization of the PCCM to an imbalanced cloner that has not been described before in the literature to the best of our knowledge. For further research, it might be interesting to try to generalize the concept of cloning machines with basis-dependent fidelities to three bases. Such a cloner could for example be applied to the 6-state protocol~\cite{sixstate} with noisy transmission lines. In the third scenario, QCL let us explicitly construct a collective attack on the BB84 protocol with a simplified version of post-processing as analyzed in Section~\ref{QCLPCCMCOLLECTIVE}, which clearly outperforms optimal individual attacks in the same setting.

Going beyond the BB84 protocol, QCL techniques can also be applied to other QKD protocols. We already mentioned the six state protocol~\cite{sixstate} in this context, but the E91 protocol~\cite{E91} is also interesting, as the $S$ value, which tests Bell's inequality, adds a constraint to the optimization. These are just two well known examples, though others could be considered. It also might be interesting to use other post-processing methods, such as larger linear codes or other reconciliation methods, against collective or coherent attacks.

On a more general note, we think that our results highlight how powerful and efficient even simple QML constructions can be.

\begin{acknowledgments}
This article was written as part of the Qu-Gov project, which was commissioned by the German Federal Ministry of Finance. We thank the Bundesdruckerei - Innovation Leadership and Team for their support and encouragement. We also thank the team at JoS for their assistance. A special thank you goes out to Prof. Jan-Åke Larsson for his valuable and insightful comments.
\end{acknowledgments}

\bibliography{ref}

\begin{thebibliography}{24}%
\makeatletter
\providecommand \@ifxundefined [1]{%
 \@ifx{#1\undefined}
}%
\providecommand \@ifnum [1]{%
 \ifnum #1\expandafter \@firstoftwo
 \else \expandafter \@secondoftwo
 \fi
}%
\providecommand \@ifx [1]{%
 \ifx #1\expandafter \@firstoftwo
 \else \expandafter \@secondoftwo
 \fi
}%
\providecommand \natexlab [1]{#1}%
\providecommand \enquote  [1]{``#1''}%
\providecommand \bibnamefont  [1]{#1}%
\providecommand \bibfnamefont [1]{#1}%
\providecommand \citenamefont [1]{#1}%
\providecommand \href@noop [0]{\@secondoftwo}%
\providecommand \href [0]{\begingroup \@sanitize@url \@href}%
\providecommand \@href[1]{\@@startlink{#1}\@@href}%
\providecommand \@@href[1]{\endgroup#1\@@endlink}%
\providecommand \@sanitize@url [0]{\catcode `\\12\catcode `\$12\catcode
  `\&12\catcode `\#12\catcode `\^12\catcode `\_12\catcode `\%12\relax}%
\providecommand \@@startlink[1]{}%
\providecommand \@@endlink[0]{}%
\providecommand \url  [0]{\begingroup\@sanitize@url \@url }%
\providecommand \@url [1]{\endgroup\@href {#1}{\urlprefix }}%
\providecommand \urlprefix  [0]{URL }%
\providecommand \Eprint [0]{\href }%
\providecommand \doibase [0]{https://doi.org/}%
\providecommand \selectlanguage [0]{\@gobble}%
\providecommand \bibinfo  [0]{\@secondoftwo}%
\providecommand \bibfield  [0]{\@secondoftwo}%
\providecommand \translation [1]{[#1]}%
\providecommand \BibitemOpen [0]{}%
\providecommand \bibitemStop [0]{}%
\providecommand \bibitemNoStop [0]{.\EOS\space}%
\providecommand \EOS [0]{\spacefactor3000\relax}%
\providecommand \BibitemShut  [1]{\csname bibitem#1\endcsname}%
\let\auto@bib@innerbib\@empty
\bibitem [{\citenamefont {Biamonte}\ \emph {et~al.}(2017)\citenamefont
  {Biamonte}, \citenamefont {Wittek}, \citenamefont {Pancotti}, \citenamefont
  {Rebentrost}, \citenamefont {Wiebe},\ and\ \citenamefont {Lloyd}}]{qml1}%
  \BibitemOpen
  \bibfield  {author} {\bibinfo {author} {\bibfnamefont {J.}~\bibnamefont
  {Biamonte}}, \bibinfo {author} {\bibfnamefont {P.}~\bibnamefont {Wittek}},
  \bibinfo {author} {\bibfnamefont {N.}~\bibnamefont {Pancotti}}, \bibinfo
  {author} {\bibfnamefont {P.}~\bibnamefont {Rebentrost}}, \bibinfo {author}
  {\bibfnamefont {N.}~\bibnamefont {Wiebe}},\ and\ \bibinfo {author}
  {\bibfnamefont {S.}~\bibnamefont {Lloyd}},\ }\bibfield  {title} {\bibinfo
  {title} {Quantum machine learning},\ }\href
  {https://doi.org/10.1038/nature23474} {\bibfield  {journal} {\bibinfo
  {journal} {Nature}\ }\textbf {\bibinfo {volume} {549}},\ \bibinfo {pages}
  {195} (\bibinfo {year} {2017})}\BibitemShut {NoStop}%
\bibitem [{\citenamefont {Schuld}\ and\ \citenamefont
  {Petruccione}(2021)}]{qml2}%
  \BibitemOpen
  \bibfield  {author} {\bibinfo {author} {\bibfnamefont {M.}~\bibnamefont
  {Schuld}}\ and\ \bibinfo {author} {\bibfnamefont {F.}~\bibnamefont
  {Petruccione}},\ }\href@noop {} {\emph {\bibinfo {title} {Machine Learning
  with Quantum Computers}}}\ (\bibinfo  {publisher} {Springer},\ \bibinfo
  {year} {2021})\BibitemShut {NoStop}%
\bibitem [{\citenamefont {Bowles}\ \emph {et~al.}(2024)\citenamefont {Bowles},
  \citenamefont {Ahmed},\ and\ \citenamefont {Schuld}}]{qmlReview}%
  \BibitemOpen
  \bibfield  {author} {\bibinfo {author} {\bibfnamefont {J.}~\bibnamefont
  {Bowles}}, \bibinfo {author} {\bibfnamefont {S.}~\bibnamefont {Ahmed}},\ and\
  \bibinfo {author} {\bibfnamefont {M.}~\bibnamefont {Schuld}},\ }\href
  {https://arxiv.org/abs/2403.07059} {\bibinfo {title} {Better than classical?
  the subtle art of benchmarking quantum machine learning models}} (\bibinfo
  {year} {2024}),\ \Eprint {https://arxiv.org/abs/2403.07059} {arXiv:2403.07059
  [quant-ph]} \BibitemShut {NoStop}%
\bibitem [{\citenamefont {{Bennett, Charles H}}\ \emph
  {et~al.}(1984)\citenamefont {{Bennett, Charles H}}, \citenamefont {{Brassard,
  Gilles}} \emph {et~al.}}]{BB84}%
  \BibitemOpen
  \bibfield  {author} {\bibinfo {author} {\bibnamefont {{Bennett, Charles H}}},
  \bibinfo {author} {\bibnamefont {{Brassard, Gilles}}}, \emph {et~al.},\
  }\href@noop {} {\bibinfo {title} {Proceedings of the ieee international
  conference on computers, systems and signal processing}} (\bibinfo {year}
  {1984})\BibitemShut {NoStop}%
\bibitem [{\citenamefont {Ekert}(1991)}]{E91}%
  \BibitemOpen
  \bibfield  {author} {\bibinfo {author} {\bibfnamefont {A.~K.}\ \bibnamefont
  {Ekert}},\ }\bibfield  {title} {\bibinfo {title} {{Quantum cryptography based
  on Bell's theorem}},\ }\href {https://doi.org/10.1103/PhysRevLett.67.661}
  {\bibfield  {journal} {\bibinfo  {journal} {\prl}\ }\textbf {\bibinfo
  {volume} {67}},\ \bibinfo {pages} {661} (\bibinfo {year} {1991})}\BibitemShut
  {NoStop}%
\bibitem [{\citenamefont {Bru\ss{}}(1998)}]{sixstate}%
  \BibitemOpen
  \bibfield  {author} {\bibinfo {author} {\bibfnamefont {D.}~\bibnamefont
  {Bru\ss{}}},\ }\bibfield  {title} {\bibinfo {title} {Optimal eavesdropping in
  quantum cryptography with six states},\ }\href
  {https://doi.org/10.1103/PhysRevLett.81.3018} {\bibfield  {journal} {\bibinfo
   {journal} {Phys. Rev. Lett.}\ }\textbf {\bibinfo {volume} {81}},\ \bibinfo
  {pages} {3018} (\bibinfo {year} {1998})}\BibitemShut {NoStop}%
\bibitem [{\citenamefont {{Shor, Peter W.}}\ and\ \citenamefont {{Preskill,
  John}}(2000)}]{Shor_2000}%
  \BibitemOpen
  \bibfield  {author} {\bibinfo {author} {\bibnamefont {{Shor, Peter W.}}}\
  and\ \bibinfo {author} {\bibnamefont {{Preskill, John}}},\ }\bibfield
  {title} {\bibinfo {title} {Simple proof of security of the bb84 quantum key
  distribution protocol},\ }\href {https://doi.org/10.1103/physrevlett.85.441}
  {\bibfield  {journal} {\bibinfo  {journal} {Physical Review Letters}\
  }\textbf {\bibinfo {volume} {85}},\ \bibinfo {pages} {441} (\bibinfo {year}
  {2000})}\BibitemShut {NoStop}%
\bibitem [{\citenamefont {Renner}(2008)}]{renner2008security}%
  \BibitemOpen
  \bibfield  {author} {\bibinfo {author} {\bibfnamefont {R.}~\bibnamefont
  {Renner}},\ }\bibfield  {title} {\bibinfo {title} {Security of quantum key
  distribution},\ }\href@noop {} {\bibfield  {journal} {\bibinfo  {journal}
  {International Journal of Quantum Information}\ }\textbf {\bibinfo {volume}
  {6}},\ \bibinfo {pages} {1} (\bibinfo {year} {2008})}\BibitemShut {NoStop}%
\bibitem [{\citenamefont {Brassard}\ and\ \citenamefont
  {Salvail}(1994)}]{Cascade}%
  \BibitemOpen
  \bibfield  {author} {\bibinfo {author} {\bibfnamefont {G.}~\bibnamefont
  {Brassard}}\ and\ \bibinfo {author} {\bibfnamefont {L.}~\bibnamefont
  {Salvail}},\ }\bibfield  {title} {\bibinfo {title} {Secret-key reconciliation
  by public discussion},\ }in\ \href@noop {} {\emph {\bibinfo {booktitle}
  {Advances in Cryptology --- EUROCRYPT '93}}},\ \bibinfo {editor} {edited by\
  \bibinfo {editor} {\bibfnamefont {T.}~\bibnamefont {Helleseth}}}\ (\bibinfo
  {publisher} {Springer Berlin Heidelberg},\ \bibinfo {address} {Berlin,
  Heidelberg},\ \bibinfo {year} {1994})\ pp.\ \bibinfo {pages}
  {410--423}\BibitemShut {NoStop}%
\bibitem [{\citenamefont {Pearson}(2004)}]{LDPC}%
  \BibitemOpen
  \bibfield  {author} {\bibinfo {author} {\bibfnamefont {D.}~\bibnamefont
  {Pearson}},\ }\bibfield  {title} {\bibinfo {title} {High‐speed qkd
  reconciliation using forward error correction},\ }\href
  {https://doi.org/10.1063/1.1834439} {\bibfield  {journal} {\bibinfo
  {journal} {AIP Conference Proceedings}\ }\textbf {\bibinfo {volume} {734}},\
  \bibinfo {pages} {299} (\bibinfo {year} {2004})}\BibitemShut {NoStop}%
\bibitem [{\citenamefont {Bennett}\ \emph {et~al.}(1988)\citenamefont
  {Bennett}, \citenamefont {Brassard},\ and\ \citenamefont
  {Robert}}]{PrivacyAmplification}%
  \BibitemOpen
  \bibfield  {author} {\bibinfo {author} {\bibfnamefont {C.~H.}\ \bibnamefont
  {Bennett}}, \bibinfo {author} {\bibfnamefont {G.}~\bibnamefont {Brassard}},\
  and\ \bibinfo {author} {\bibfnamefont {J.-M.}\ \bibnamefont {Robert}},\
  }\bibfield  {title} {\bibinfo {title} {Privacy amplification by public
  discussion},\ }\href {https://doi.org/10.1137/0217014} {\bibfield  {journal}
  {\bibinfo  {journal} {SIAM Journal on Computing}\ }\textbf {\bibinfo {volume}
  {17}},\ \bibinfo {pages} {210} (\bibinfo {year} {1988})}\BibitemShut
  {NoStop}%
\bibitem [{\citenamefont {{Pirandola, S.}}\ \emph {et~al.}(2020)\citenamefont
  {{Pirandola, S.}}, \citenamefont {{Andersen, U. L.}}, \citenamefont {{Banchi,
  L.}}, \citenamefont {{Berta, M.}}, \citenamefont {{Bunandar, D.}},
  \citenamefont {{Colbeck, R.}}, \citenamefont {{Englund, D.}}, \citenamefont
  {{Gehring, T.}}, \citenamefont {{Lupo, C.}}, \citenamefont {{Ottaviani, C.}},
  \citenamefont {{Pereira, J. L.}}, \citenamefont {{Razavi, M.}}, \citenamefont
  {{Shamsul Shaari, J.}}, \citenamefont {{Tomamichel, M.}}, \citenamefont
  {{Usenko, V. C.}}, \citenamefont {{Vallone, G.}}, \citenamefont {{Villoresi,
  P.}},\ and\ \citenamefont {{Wallden, P.}}}]{bigReview}%
  \BibitemOpen
  \bibfield  {author} {\bibinfo {author} {\bibnamefont {{Pirandola, S.}}},
  \bibinfo {author} {\bibnamefont {{Andersen, U. L.}}}, \bibinfo {author}
  {\bibnamefont {{Banchi, L.}}}, \bibinfo {author} {\bibnamefont {{Berta,
  M.}}}, \bibinfo {author} {\bibnamefont {{Bunandar, D.}}}, \bibinfo {author}
  {\bibnamefont {{Colbeck, R.}}}, \bibinfo {author} {\bibnamefont {{Englund,
  D.}}}, \bibinfo {author} {\bibnamefont {{Gehring, T.}}}, \bibinfo {author}
  {\bibnamefont {{Lupo, C.}}}, \bibinfo {author} {\bibnamefont {{Ottaviani,
  C.}}}, \bibinfo {author} {\bibnamefont {{Pereira, J. L.}}}, \bibinfo {author}
  {\bibnamefont {{Razavi, M.}}}, \bibinfo {author} {\bibnamefont {{Shamsul
  Shaari, J.}}}, \bibinfo {author} {\bibnamefont {{Tomamichel, M.}}}, \bibinfo
  {author} {\bibnamefont {{Usenko, V. C.}}}, \bibinfo {author} {\bibnamefont
  {{Vallone, G.}}}, \bibinfo {author} {\bibnamefont {{Villoresi, P.}}},\ and\
  \bibinfo {author} {\bibnamefont {{Wallden, P.}}},\ }\bibfield  {title}
  {\bibinfo {title} {Advances in quantum cryptography},\ }\href
  {https://doi.org/10.1364/aop.361502} {\bibfield  {journal} {\bibinfo
  {journal} {Advances in Optics and Photonics}\ }\textbf {\bibinfo {volume}
  {12}},\ \bibinfo {pages} {1012} (\bibinfo {year} {2020})}\BibitemShut
  {NoStop}%
\bibitem [{\citenamefont {Mitarai}\ \emph {et~al.}(2018)\citenamefont
  {Mitarai}, \citenamefont {Negoro}, \citenamefont {Kitagawa},\ and\
  \citenamefont {Fujii}}]{QCL}%
  \BibitemOpen
  \bibfield  {author} {\bibinfo {author} {\bibfnamefont {K.}~\bibnamefont
  {Mitarai}}, \bibinfo {author} {\bibfnamefont {M.}~\bibnamefont {Negoro}},
  \bibinfo {author} {\bibfnamefont {M.}~\bibnamefont {Kitagawa}},\ and\
  \bibinfo {author} {\bibfnamefont {K.}~\bibnamefont {Fujii}},\ }\bibfield
  {title} {\bibinfo {title} {Quantum circuit learning},\ }\bibfield  {journal}
  {\bibinfo  {journal} {Physical Review A}\ }\textbf {\bibinfo {volume} {98}},\
  \href {https://doi.org/10.1103/physreva.98.032309}
  {10.1103/physreva.98.032309} (\bibinfo {year} {2018})\BibitemShut {NoStop}%
\bibitem [{\citenamefont {Bru\ss{}}\ \emph {et~al.}(2000)\citenamefont
  {Bru\ss{}}, \citenamefont {Cinchetti}, \citenamefont {Mauro~D'Ariano},\ and\
  \citenamefont {Macchiavello}}]{PCCM}%
  \BibitemOpen
  \bibfield  {author} {\bibinfo {author} {\bibfnamefont {D.}~\bibnamefont
  {Bru\ss{}}}, \bibinfo {author} {\bibfnamefont {M.}~\bibnamefont {Cinchetti}},
  \bibinfo {author} {\bibfnamefont {G.}~\bibnamefont {Mauro~D'Ariano}},\ and\
  \bibinfo {author} {\bibfnamefont {C.}~\bibnamefont {Macchiavello}},\
  }\bibfield  {title} {\bibinfo {title} {Phase-covariant quantum cloning},\
  }\href {https://doi.org/10.1103/PhysRevA.62.012302} {\bibfield  {journal}
  {\bibinfo  {journal} {Phys. Rev. A}\ }\textbf {\bibinfo {volume} {62}},\
  \bibinfo {pages} {012302} (\bibinfo {year} {2000})}\BibitemShut {NoStop}%
\bibitem [{\citenamefont {Coyle}\ \emph {et~al.}(2022)\citenamefont {Coyle},
  \citenamefont {Doosti}, \citenamefont {Kashefi},\ and\ \citenamefont
  {Kumar}}]{VQC}%
  \BibitemOpen
  \bibfield  {author} {\bibinfo {author} {\bibfnamefont {B.}~\bibnamefont
  {Coyle}}, \bibinfo {author} {\bibfnamefont {M.}~\bibnamefont {Doosti}},
  \bibinfo {author} {\bibfnamefont {E.}~\bibnamefont {Kashefi}},\ and\ \bibinfo
  {author} {\bibfnamefont {N.}~\bibnamefont {Kumar}},\ }\bibfield  {title}
  {\bibinfo {title} {Progress toward practical quantum cryptanalysis by
  variational quantum cloning},\ }\bibfield  {journal} {\bibinfo  {journal}
  {Physical Review A}\ }\textbf {\bibinfo {volume} {105}},\ \href
  {https://doi.org/10.1103/physreva.105.042604} {10.1103/physreva.105.042604}
  (\bibinfo {year} {2022})\BibitemShut {NoStop}%
\bibitem [{\citenamefont {Molotkov}\ and\ \citenamefont
  {Timofeev}(2007)}]{molotkov2007explicit}%
  \BibitemOpen
  \bibfield  {author} {\bibinfo {author} {\bibfnamefont {S.}~\bibnamefont
  {Molotkov}}\ and\ \bibinfo {author} {\bibfnamefont {A.}~\bibnamefont
  {Timofeev}},\ }\bibfield  {title} {\bibinfo {title} {Explicit attack on the
  key in quantum cryptography (bb84 protocol) reaching the theoretical error
  limit {$Q_c\approx 11\%$}},\ }\href@noop {} {\bibfield  {journal} {\bibinfo
  {journal} {JETP Letters}\ }\textbf {\bibinfo {volume} {85}},\ \bibinfo
  {pages} {524} (\bibinfo {year} {2007})}\BibitemShut {NoStop}%
\bibitem [{\citenamefont {{Cerezo, M.}}\ \emph {et~al.}(2021)\citenamefont
  {{Cerezo, M.}}, \citenamefont {{Arrasmith, Andrew}}, \citenamefont {{Babbush,
  Ryan}}, \citenamefont {{Benjamin, Simon C.}}, \citenamefont {{Endo, Suguru}},
  \citenamefont {{Fujii, Keisuke}}, \citenamefont {{McClean, Jarrod R.}},
  \citenamefont {{Mitarai, Kosuke}}, \citenamefont {{Yuan, Xiao}},
  \citenamefont {{Cincio, Lukasz}},\ and\ \citenamefont {{Coles, Patrick
  J.}}}]{VQA}%
  \BibitemOpen
  \bibfield  {author} {\bibinfo {author} {\bibnamefont {{Cerezo, M.}}},
  \bibinfo {author} {\bibnamefont {{Arrasmith, Andrew}}}, \bibinfo {author}
  {\bibnamefont {{Babbush, Ryan}}}, \bibinfo {author} {\bibnamefont {{Benjamin,
  Simon C.}}}, \bibinfo {author} {\bibnamefont {{Endo, Suguru}}}, \bibinfo
  {author} {\bibnamefont {{Fujii, Keisuke}}}, \bibinfo {author} {\bibnamefont
  {{McClean, Jarrod R.}}}, \bibinfo {author} {\bibnamefont {{Mitarai,
  Kosuke}}}, \bibinfo {author} {\bibnamefont {{Yuan, Xiao}}}, \bibinfo {author}
  {\bibnamefont {{Cincio, Lukasz}}},\ and\ \bibinfo {author} {\bibnamefont
  {{Coles, Patrick J.}}},\ }\bibfield  {title} {\bibinfo {title} {Variational
  quantum algorithms},\ }\href {https://doi.org/10.1038/s42254-021-00348-9}
  {\bibfield  {journal} {\bibinfo  {journal} {Nature Reviews Physics}\ }\textbf
  {\bibinfo {volume} {3}},\ \bibinfo {pages} {625} (\bibinfo {year}
  {2021})}\BibitemShut {NoStop}%
\bibitem [{\citenamefont {Kandala}\ \emph {et~al.}(2017)\citenamefont
  {Kandala}, \citenamefont {Mezzacapo}, \citenamefont {Temme}, \citenamefont
  {Takita}, \citenamefont {Brink}, \citenamefont {Chow},\ and\ \citenamefont
  {Gambetta}}]{HEA}%
  \BibitemOpen
  \bibfield  {author} {\bibinfo {author} {\bibfnamefont {A.}~\bibnamefont
  {Kandala}}, \bibinfo {author} {\bibfnamefont {A.}~\bibnamefont {Mezzacapo}},
  \bibinfo {author} {\bibfnamefont {K.}~\bibnamefont {Temme}}, \bibinfo
  {author} {\bibfnamefont {M.}~\bibnamefont {Takita}}, \bibinfo {author}
  {\bibfnamefont {M.}~\bibnamefont {Brink}}, \bibinfo {author} {\bibfnamefont
  {J.~M.}\ \bibnamefont {Chow}},\ and\ \bibinfo {author} {\bibfnamefont
  {J.~M.}\ \bibnamefont {Gambetta}},\ }\bibfield  {title} {\bibinfo {title}
  {Hardware-efficient variational quantum eigensolver for small molecules and
  quantum magnets},\ }\href {https://doi.org/10.1038/nature23879} {\bibfield
  {journal} {\bibinfo  {journal} {Nature}\ }\textbf {\bibinfo {volume} {549}},\
  \bibinfo {pages} {242} (\bibinfo {year} {2017})}\BibitemShut {NoStop}%
\bibitem [{\citenamefont {{Schuld, Maria}}\ \emph {et~al.}(2020)\citenamefont
  {{Schuld, Maria}}, \citenamefont {{Bocharov, Alex}}, \citenamefont {{Svore,
  Krysta M.}},\ and\ \citenamefont {{Wiebe, Nathan}}}]{HEA_schuld}%
  \BibitemOpen
  \bibfield  {author} {\bibinfo {author} {\bibnamefont {{Schuld, Maria}}},
  \bibinfo {author} {\bibnamefont {{Bocharov, Alex}}}, \bibinfo {author}
  {\bibnamefont {{Svore, Krysta M.}}},\ and\ \bibinfo {author} {\bibnamefont
  {{Wiebe, Nathan}}},\ }\bibfield  {title} {\bibinfo {title} {Circuit-centric
  quantum classifiers},\ }\bibfield  {journal} {\bibinfo  {journal} {Physical
  Review A}\ }\textbf {\bibinfo {volume} {101}},\ \href
  {https://doi.org/10.1103/physreva.101.032308} {10.1103/physreva.101.032308}
  (\bibinfo {year} {2020})\BibitemShut {NoStop}%
\bibitem [{\citenamefont {Kingma}\ and\ \citenamefont
  {Ba}(2015)}]{AdamOptimizer}%
  \BibitemOpen
  \bibfield  {author} {\bibinfo {author} {\bibfnamefont {D.}~\bibnamefont
  {Kingma}}\ and\ \bibinfo {author} {\bibfnamefont {J.}~\bibnamefont {Ba}},\
  }\bibfield  {title} {\bibinfo {title} {Adam: A method for stochastic
  optimization},\ }in\ \href@noop {} {\emph {\bibinfo {booktitle}
  {International Conference on Learning Representations (ICLR)}}}\ (\bibinfo
  {address} {San Diega, CA, USA},\ \bibinfo {year} {2015})\BibitemShut
  {NoStop}%
\bibitem [{\citenamefont {et~al.}(2022)}]{Pennylane}%
  \BibitemOpen
  \bibfield  {author} {\bibinfo {author} {\bibfnamefont {V.~B.}\ \bibnamefont
  {et~al.}},\ }\href@noop {} {\bibinfo {title} {Pennylane: Automatic
  differentiation of hybrid quantum-classical computations}} (\bibinfo {year}
  {2022}),\ \Eprint {https://arxiv.org/abs/1811.04968} {arXiv:1811.04968
  [quant-ph]} \BibitemShut {NoStop}%
\bibitem [{\citenamefont {Renner}\ \emph {et~al.}(2005)\citenamefont {Renner},
  \citenamefont {Gisin},\ and\ \citenamefont {Kraus}}]{infoSecurityProof}%
  \BibitemOpen
  \bibfield  {author} {\bibinfo {author} {\bibfnamefont {R.}~\bibnamefont
  {Renner}}, \bibinfo {author} {\bibfnamefont {N.}~\bibnamefont {Gisin}},\ and\
  \bibinfo {author} {\bibfnamefont {B.}~\bibnamefont {Kraus}},\ }\bibfield
  {title} {\bibinfo {title} {Information-theoretic security proof for
  quantum-key-distribution protocols},\ }\bibfield  {journal} {\bibinfo
  {journal} {Physical Review A}\ }\textbf {\bibinfo {volume} {72}},\ \href
  {https://doi.org/10.1103/physreva.72.012332} {10.1103/physreva.72.012332}
  (\bibinfo {year} {2005})\BibitemShut {NoStop}%
\bibitem [{\citenamefont {Hausladen}\ and\ \citenamefont
  {Wootters}(1994)}]{PGM}%
  \BibitemOpen
  \bibfield  {author} {\bibinfo {author} {\bibfnamefont {P.}~\bibnamefont
  {Hausladen}}\ and\ \bibinfo {author} {\bibfnamefont {W.~K.}\ \bibnamefont
  {Wootters}},\ }\bibfield  {title} {\bibinfo {title} {A 'pretty
  good'measurement for distinguishing quantum states},\ }\href@noop {}
  {\bibfield  {journal} {\bibinfo  {journal} {Journal of Modern Optics}\
  }\textbf {\bibinfo {volume} {41}},\ \bibinfo {pages} {2385} (\bibinfo {year}
  {1994})}\BibitemShut {NoStop}%
\bibitem [{\citenamefont {Helstrom}(1969)}]{helstrom1969quantum}%
  \BibitemOpen
  \bibfield  {author} {\bibinfo {author} {\bibfnamefont {C.~W.}\ \bibnamefont
  {Helstrom}},\ }\bibfield  {title} {\bibinfo {title} {Quantum detection and
  estimation theory},\ }\href@noop {} {\bibfield  {journal} {\bibinfo
  {journal} {Journal of Statistical Physics}\ }\textbf {\bibinfo {volume}
  {1}},\ \bibinfo {pages} {231} (\bibinfo {year} {1969})}\BibitemShut {NoStop}%
\end{thebibliography}%

\end{document}